# Hiding Data in OFDM Symbols of IEEE 802.11 Networks

Krzysztof Szczypiorski, Wojciech Mazurczyk
Warsaw University of Technology, Institute of Telecommunications
ul. Nowowiejska 15/19, 00-665 Warsaw, Poland
E-mail: {ksz, wmazurczyk}@tele.pw.edu.pl

*Abstract* — **This paper presents a new steganographic method called WiPad (Wireless Padding). It is based on the insertion of hidden data into the padding of frames at the physical layer of WLANs (*Wireless Local Area Networks*). A performance analysis based on a Markov model, previously introduced and validated by the authors in [10], is provided for the method in relation to the IEEE 802.11 a/g standards . Its results prove that maximum steganographic bandwidth for WiPad is as high as 1.1 Mbit/s for data frames and 0.44 Mbit/s for acknowledgment (ACK) frames. To the authors' best knowledge this is the most capacious of all the known steganographic network channels.**

## I. INTRODUCTION

In this paper we present and evaluate a new information hiding method based on bit padding of OFDM (*Orthogonal Frequency Division Multiplexing*) symbols at the physical layer (PHY) of IEEE 802.11 networks. Depending on the transmission data rate at the PHY layer the number of encoded bits per symbol spans from 24 up to 216, therefore as many as 27 octets can be embedded in each OFDM symbol. Due to the specific structure of a frame (described in detail in Section III) up to 210 bits per frame (26 ¼ octets/frame) can be allocated for hidden communication. We named this steganographic method utilizing the principle of frame padding in the physical layer of WLANs with the acronym WiPad (Wireless Padding).

This paper provides an evaluation of throuput for this method with the aid of our general, Markov-based model introduced and validated in [10]. This model is in line with the extensions of Bianchi's basic model [2] proposed in [10] and [9]. The essential difference with respect to the latter two is the consideration of the effect of freezing of the stations' backoff timer, as well as the limitation of the number of retransmissions and the maximum size of the contention window, and the impact of transmission errors. Results presented in [10] proved good accuracy of our model in the case of both: error-free and error-prone channels. In either case the proposed model is more accurate than other models presented in literature with which it was compared (including: [1], [9] and [10]), most notably, when large numbers of stations are under consideration.

This paper is organized as follows. Next section provides an overview of the state of the art with regard to information hiding techniques that utilize padding in WLANs. Section III contains a description of our method. Section IV is a brief overview of the model presented in [10] and introduces a performance metric for the proposed method. Section V presents a performance analysis of the method based on the given model. Finally, Section VI contains conclusions and suggestions for future work.

## II. STATE OF THE ART

Data padding can be found at any layer of the OSI RM, but it is typically exploited for covert communications only in the data link, network and transport layers. Wolf proposed in [13] a steganographic method utilizing padding of 802.3 frames. Its achievable steganographic capacity was maximally 45 bytes/frame. Fisk et al. [2] presented padding of the IP and TCP headers in the context of active wardens. Each of these fields offers up to 31 bits/packet for covert communication. Jankowski et al. in [4] developed a steganographic system, PadSteg, which bases on Ethernet frames' padding and is used in conjunction with ARP (*Address Resolution Protocol*) and TCP (*Transmission Control Protocol*) protocols. Padding of IPv6 packets as means for information hiding was described by Lucena et al. in [8] - it offers a couple of channels with a steganographic bandwidth reaching 256 bytes/packet.

Steganography for IEEE 802.11 was proposed by Szczypiorski in [12], who postulated usage of frames with intentionally corrupted checksums to establish covert communication. The system was evaluated by Szczypiorski in [11]. Krätzer et al. in [6] developed a storage channel based scenario (employing header embedding) and a time channel based scenario for IEEE 802.11. In [7] Krätzer et al. reconsidered the approach presented in [6].

## III. THE METHOD

IEEE 802.11 a/g standards exploit OFDM at the physical layer. 802.11 network's PHY layer consists of two sublayers: PLCP (*PHY Layer Convergance Procedure*) and PHY Medium-Dependent (PMD). Selection of a specific transmission data rate at the PHY layer implies functioning with a predefined number of bits corresponding to each OFDM symbol. The number of bits per symbol may vary from 24, for 6 Mbps, up to 216, for 54 Mbps (Table I). Three fields are liable to padding: SERVICE, PSDU (*Physical layer Service Data Unit*), TAIL (Fig. 1). The lengths of SERVICE and TAIL are constant (16 bits and 6 bits respectively), while the PSDU is a MAC frame and its length varies depending on user data, ciphers and network operation mode (ad hoc vs. infrastructure).

TABLE I PARAMETERS OF 802.11 A/G OFDM PHY

| Rate $R$ [Mbit/s] | Modulation | Code rate | Number of bits per symbol – $N_{BpS}$ | Factorization of $N_{BpS}$ into primes |
|---|---|---|---|---|
| 6 | BPSK | ½ | 24 | $2^3 3$ |
| 9 | BPSK | ¾ | 36 | $2^2 3^2$ |
| 12 | QPSK | ½ | 48 | $2^4 3$ |
| 18 | QPSK | ¾ | 72 | $2^3 3^2$ |
| 24 | 16-QAM | ½ | 96 | $2^5 3$ |
| 36 | 16-QAM | ¾ | 144 | $2^4 3^2$ |
| 48 | 64-QAM | ⅔ | 192 | $2^6 3$ |
| 54 | 64-QAM | ¾ | 216 | $2^3 3^3$ |

For each rate R, the number of bits per symbol can be factorized into primes (Table I) and then, using this knowledge, a least common multiple can be calculated as $2^6 3^3 = 1728$. This means that the maximum number of padding bytes (octets) that may be used for all rates is:

$$L_\alpha = \frac{2^6 3^3}{8}\alpha - 2 = 216\alpha - 2 \quad (1)$$

where $\alpha$ is a positive integer.

Please note that padding is present in all frames, therefore frames that are more frequently exchanged, like ACKs may become an interesting target for covert communication.

Typically all padding bits are set to zero [3], but in this paper we assuume that all of them could be used for steganographic purposes.

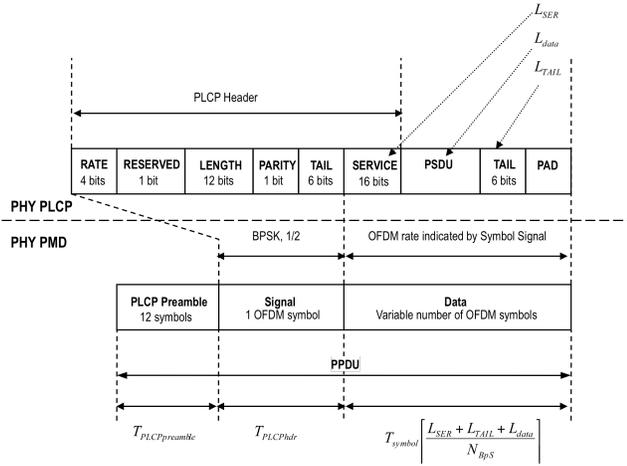

Fig. 1. The structure of 802.11a/g PPDU for ERP-OFDM networks

IV. THE MODEL

### A. Assumptions

We considered saturation conditions: stations have non empty queues and there is always a frame to be sent. The number of stations competing for medium access is *n* (for *n*=1 there is one station sending frames to another station which may only reply with an ACK frame). Errors in the transmission medium are fully randomly distributed; this is the worst case scenario in terms of *frame error rate* – FER. All stations experience the same *bit error rate* (BER) and all are within each other's transmission range and there are no hidden terminals. Stations communicate in ad hoc mode (BSS – *Basic Service Set*) with basic access method. Every station employs the same physical layer (PHY). The transmission data rate *R* is the same and constant for all stations. All frames are of constant length *L*. The only frames that are exchanged are data frames and ACK frames. Collided frames are discarded – the capture effect [5] is not considered.

### B. Saturation throughput S expressed through characteristics of the physical channel

The saturation throughput *S* is defined as in [1]:

$$S = \frac{E[DATA]}{E[T]} \quad (2)$$

where E[*DATA*] is the mean value of successfully transmitted payload, and E[*T*] is the mean value of the duration of the following *channel states*:
- $T_I$ – idle slot,
- $T_S$ – successful transmission,
- $T_C$ – transmission with collision,
- $T_{E\_DATA}$ – unsuccessful transmission with data frame error,
- $T_{E\_ACK}$ – unsuccessful transmission with ACK error.

Fig. 2 illustrates dependence of the above channel states on:
- $T_{PHYhdr}$ – duration of a PLCP (*PHY Layer Convergence Procedure*) preamble and a PLCP header,
- $T_{DATA}$ – data frame transmission time,
- $T_{ACK}$ – ACK frame duration,
- $T_{SIFS}$ – duration of SIFS (*Short InterFrame Space*),
- $T_{DIFS}$ – duration of DIFS (*DCF InterFrame Space*),
- $T_{EIFS}$ – duration of EIFS (*Extended InterFrame Space*).

The relation of the saturation throughput to physical channel characteristics is calculated similarly as in [9]:

$$\begin{cases} T_I = \sigma \\ T_S = 2T_{PHYhdr} + T_{DATA} + 2\delta + T_{SIFS} + T_{ACK} + T_{DIFS} \\ T_C = T_{PHYhdr} + T_{DATA} + \delta + T_{EIFS} \\ T_{E\_DATA} = T_{PHYhdr} + \delta + T_{DATA} + T_{EIFS} \\ T_{E\_ACK} = T_S \end{cases} \quad (3)$$

where $\sigma$ is the duration of an idle slot (*aSlotTime* [3]) and $\delta$ is the propagation delay.

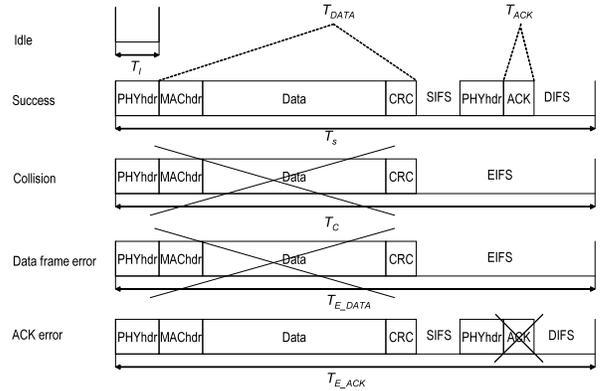

Fig. 2. Channel states

For 802.11a/g OFDM PHY (Fig. 1):

$$T_{ACK} = T_{symbol}\left\lceil\frac{L_{SER} + L_{TAIL} + L_{ACK}}{N_{BpS}}\right\rceil \quad (4)$$

$$T_{DATA} = T_{symbol}\left\lceil\frac{L_{SER} + L_{TAIL} + L_{DATA}}{N_{BpS}}\right\rceil \quad (5)$$

where:
- $T_{symbol}$ – duration of a transmission symbol,
- $L_{SER}$ – OFDM PHY layer SERVICE field size,
- $L_{TAIL}$ – OFDM PHY layer TAIL field size,
- $N_{BpS}$ – number of encoded bits per symbol,

$L_{ACK}$ – size of an ACK frame,
$L_{DATA}$ – size of a data frame.

Values of $\sigma$, $T_{PHYhdr}$, $T_{SIFS}$, $T_{DIFS}$, $T_{EIFS}$, $T_{symbol}$, $N_{BpS}$, $L_{SER}$ and $L_{TAIL}$ are defined in accordance with the 802.11 standard ([3]).

Probabilities corresponding to the states of the channel are denoted as follows:

$P_I$ – probability of an idle slot,
$P_S$ – probability of successful transmission,
$P_C$ – probability of collision,
$P_{E\_DATA}$ – probability of unsuccessful transmission due to data frame error,
$P_{E\_ACK}$ – probability of unsuccessful transmission due to ACK error.

Let $\tau$ be the probability of frame transmission, $p_{e\_data}$ the probability of data frame error and $p_{e\_ACK}$ the probability of an ACK error. These are related to channel state probabilities as follows:

$$\begin{cases} P_I = (1-\tau)^n \\ P_S = n\tau(1-\tau)^{n-1}(1-p_{e\_data})(1-p_{e\_ACK}) \\ P_C = 1-(1-\tau)^n - n\tau(1-\tau)^{n-1} \\ P_{E\_DATA} = n\tau(1-\tau)^{n-1}p_{e\_data} \\ P_{E\_ACK} = n\tau(1-\tau)^{n-1}(1-p_{e\_data})p_{e\_ACK} \end{cases} \quad (6)$$

The saturation throughput $S$ equals:

$$S = \frac{P_S L_{pld}}{T_I P_I + T_S P_S + T_C P_C + T_{E\_DATA} P_{E\_DATA} + T_{E\_ACK} P_{E\_ACK}} \quad (7)$$

where $L_{pld}$ is MAC (*Medium Access Control*) payload size and $L_{pld} = L - L_{MAChdr}$, where $L_{MAChdr}$ is the size of the MAC header plus the size of a FCS (*Frame Checksum Sequence*).

The data rate $R$ is defined as:

$$R = \frac{N_{BpS}}{T_{symbol}} \quad (8)$$

As a result, saturation throughput $S$ is expressed as a function of $\tau$, $p_{e\_data}$ and $p_{e\_ACK}$. In the following sections these probabilities are evaluated.

## C. Probability of frame transmission $\tau$

Let $s(t)$ be a random variable describing DCF backoff stage at time $t$, with values from set $\{0, 1, 2,…,m\}$. Let $b(t)$ be a random variable describing the value of the backoff timer at time $t$, with values from the set $\{0, 1, 2,…, W_i-1\}$. These random variables are correlated because the maximum value of the backoff timer depends on the backoff stage:

$$W_i = \begin{cases} 2^i W_0, & i \leq m' \\ 2^{m'} W_0 = W_m, & i > m' \end{cases} \quad (9)$$

where $W_0$ is the initial size of the contention window (*CW*) and $m'$ is (the boundary stage above which the contention widow will not be enlarged further); $m'$ can be either greater, smaller or $m$. $W_0$ and $W_{m'}$ depend on $CW_{min}$ and $CW_{max}$ [3]:

$$W_0 = CW_{min} + 1 \quad (10)$$

$$W_{m'} = CW_{max} + 1 = 2^{m'} W_0 \quad (11)$$

The two-dimensional process $(s(t), b(t))$ will be analyzed with the aid of an embedded Markov chain (steady state probabilities), whose states correspond to the time instants at which the channel state changes. Let $(i,k)$ denote the current state of this process. The conditional, one-step, state transition probabilities will be denoted by $P = (\cdot,\cdot|\cdot,\cdot)$.

Let $p_f$ be the probability of transmission failure and $p_{coll}$ the probability of collision. The non-null transition probabilities are determined as follows:

$$\begin{aligned} &(a)\, P(i,k|i,k+1) = 1 - p_{coll}, && 0 \leq i \leq m, 0 \leq k \leq W_i - 2 \\ &(b)\, P(i,k|i,k) = p_{coll}, && 0 \leq i \leq m, 1 \leq k \leq W_i - 1 \\ &(c)\, P(0,k|i,0) = (1-p_f)/W_0, && 0 \leq i \leq m-1, 0 \leq k \leq W_0 - 1 \\ &(d)\, P(i,k|i-1,0) = p_f/W_i, && 1 \leq i \leq m, 0 \leq k \leq W_i - 1 \\ &(e)\, P(0,k|m,0) = 1/W_0, && 0 \leq k \leq W_0 - 1 \end{aligned} \quad (12)$$

Ad (a): The station's backoff timer is decremented from $k+1$ to $k$ at a fixed, *i-th* backoff stage, i.e. the station has detected an idle slot. The probability of this event $Pr\{channel\ is\ idle\} = 1 - Pr\{one\ or\ more\ stations\ are\ transmitting\}$. We consider saturation conditions, so $Pr\{one\ or\ more\ stations\ are\ transmitting\}$ equals $p_{coll}$.

Ad (b): The station's backoff timer is frozen at a fixed, *i-th* backoff stage, i.e. the channel is busy. $Pr\{channel\ is\ busy\} = Pr\{one\ or\ more\ stations\ are\ transmitting\} = p_{coll}$.

Ad (c): The station's backoff timer is changed from *0* to *k* and the backoff stage reinitialized from *i* to *0*. The probability of this event equals: $Pr\{transmission\ is\ successful\ and\ number\ k\ was\ randomly\ chosen\ to\ initiate\ the\ backoff\ timer\ at\ stage\ 0\} = Pr\{transmission\ is\ successful\} \cdot Pr\{number\ k\ was\ randomly\ chosen\ to\ initiate\ the\ backoff\ timer\ at\ stage\ 0\}$. The probability of successful transmission is equal to $1 - p_f$ and the probability that number *k* was randomly chosen to initiate the backoff timer at stage *0* equals $1/W_0$.

Ad (d): The station's backoff timer is changed from *0* to *k* and the backoff stage is increased from *i*-1 to *i*. Probability of this event equals: $Pr\{transmission\ is\ unsuccessful\ and\ number\ k\ was\ randomly\ chosen\ to\ initiate\ the\ backoff\ timer\ at\ stage\ i\} = Pr\{transmission\ is\ unsuccessful\} \cdot Pr\{number\ k\ was\ randomly\ chosen\ to\ initiate\ the\ backoff\ timer\ at\ stage\ i\}$. The probability of unsuccessful transmission equals $p_f$ and the probability that number *k* was randomly chosen to initiate the backoff timer at stage *i* equals $1/W_i$.

Ad (e): The station's backoff timer is changed from *0* to *k* and the backoff stage is changed from *m* to *0*, i.e. the station has reached the maximum retransmission count. The probability of this event equals the probability that number *k* was randomly chosen to initiate the backoff timer at stage *0*, i.e. $1/W_0$.

Let $b_{i,k}$ be the steady-state occupancy probability of state $(i,k)$. It can be shown that:

$$b_{i,0} = p_f \cdot b_{i-1,0} \quad (13)$$

$$b_{i,0} = p_f^i \cdot b_{0,0} \quad (14)$$

and

$$b_{i,k} = \begin{cases} \dfrac{W_i - k}{W_i(1-p_{coll})} p_f^i \cdot b_{0,0}, & 0 < k \leq W_i - 1 \\ p_f^i \cdot b_{0,0}, & k = 0 \end{cases} \quad (15)$$

From the normalization condition:

$$\sum_{i=0}^{m} \sum_{k=0}^{W_i - 1} b_{i,k} = 1 \quad (16)$$

and

$$\sum_{i=0}^{m} b_{i,0} = b_{0,0} \frac{1-p_f^{m+1}}{1-p_f} \quad (17)$$

we get:

$$b_{0,0}^{-1} = \begin{cases} \frac{(1-p_f)W_0(1-(2p_f)^{m+1})-(1-2p_f)(1-p_f^{m+1})}{2(1-2p_f)(1-p_f)(1-p_{coll})} + \frac{1-p_f^{m+1}}{1-p_f}, & m \le m' \\ \frac{\Psi}{2(1-2p_f)(1-p_f)(1-p_{coll})} + \frac{1-p_f^{m+1}}{1-p_f}, & m > m' \end{cases} \quad (18)$$

where

$$\Psi = (1-p_f)W_0(1-(2p_f)^{m'+1})-(1-2p_f)(1-p_f^{m+1}) + W_0 2^{m'} p_f^{m'+1}(1-2p_f)(1-p_f^{m-m'}) \quad (19)$$

The probability of frame transmission $\tau$ is equal to Pr{backoff timer equals 0} and thus:

$$\tau = \sum_{i=0}^{m} b_{i,0} = \begin{cases} \left(\frac{(1-p_f)W_0(1-(2p_f)^{m+1})-(1-2p_f)(1-p_f^{m+1})}{2(1-2p_f)(1-p_f)(1-p_{coll})} + \frac{1-p_f^{m+1}}{1-p_f}\right)^{-1} \frac{1-p_f^{m+1}}{1-p_f}, & m \le m' \\ \left(\frac{\Psi}{2(1-2p_f)(1-p_f)(1-p_{coll})} + \frac{1-p_f^{m+1}}{1-p_f}\right)^{-1} \frac{1-p_f^{m+1}}{1-p_f}, & m > m' \end{cases} \quad (20)$$

For $p_{coll} = 0$ the above solution is the same as presented in [9].

### D. Probability of transmission failure $p_f$ and probability of collision $p_{coll}$

The probability of transmission failure

$$p_f = 1 - (1-p_{coll})(1-p_e) \quad (21)$$

where $p_e$ is the frame error probability:

$$p_e = 1 - (1-p_{e\_data})(1-p_{e\_ACK}) \quad (22)$$

where $p_{e\_data}$ is FER for data frames and $p_{e\_ACK}$ is FER for ACK frames. $p_{e\_data}$ and $p_{e\_ACK}$ can be calculated from bit error probability (i.e. BER), $p_b$:

$$p_{e\_data} = 1 - (1-p_b)^{L_{data}} \quad (23)$$
$$p_{e\_ACK} = 1 - (1-p_b)^{L_{ACK}} \quad (24)$$

The probability of collision:

$$p_{coll} = 1 - (1-\tau)^{n-1} \quad (25)$$

Finally

$$p_f = 1 - (1-p_{coll})(1-p_e) = 1 - (1-\tau)^{n-1}(1-p_e) \quad (26)$$

Equations (20) and (26) form a non-linear system with two unknown variables $\tau$ and $p_f$ which can be solved numerically.

### E. Capacity and saturation throughput of steganographic channels

Let the capacity of a steganographic channel based on data frames be:

$$C_{DATA} = N_{BpS} \left\lceil \frac{L_{SER} + L_{TAIL} + L_{DATA}}{N_{BpS}} \right\rceil - (L_{SER} + L_{TAIL} + L_{DATA}) \quad (27)$$

Let the capacity of a steganographic channel based on ACK frames be:

$$C_{ACK} = N_{BpS} \left\lceil \frac{L_{SER} + L_{TAIL} + L_{ACK}}{N_{BpS}} \right\rceil - (L_{SER} + L_{TAIL} + L_{ACK}) \quad (28)$$

Therefore the saturation throughput of a steganographic channel based on data frames may be defined as:

$$S_{DATA} = \frac{C_{DATA} \cdot S}{n \cdot L_{pld}} \quad (29)$$

And, finally, the saturation throughput of a steganographic channel based on ACK frames equals:

$$S_{ACK} = \frac{C_{ACK} \cdot S}{n \cdot L_{pld}} \quad (30)$$

## V. ANALYSIS

### A. Frames with a maximum number of padding octets

All diagrams presented in this section display values of the saturation throughput of the proposed steganographic method (WiPad) based on the data frame variant. All calculations were made for $n \in \{1, 2, 3, 4, 5, 6, 7, 8, 9, 10\}$. For $L=214$ octet frames ($\alpha=1$; 186 bytes at IP layer) the following values of BER were used $\{10^{-4}, 10^{-5}, 0\}$, and for $L \in \{214, 430, 646, 862, 1078, 1294, 1510\}$ octet frames ($\alpha \in \{1, 2,..., 7\}$) the correspondent $BER \in \{10^{-4}, 10^{-5}, 0\}$. We considered the IEEE 802.11g – ERP-OFDM i.e. "g"-only mode and a data rate of $R=54$ Mbps (with an exception for the last diagram, which provides an evaluation of the impact of $R$ on $S_{DATA}$).

Fig. 3 presents $S_{DATA}$ as a function of $n$ for $L=214$ octet frames and different values of BER. Along with an increasing value of BER the steganographic throughput, $S_{DATA}$, declines. The maximum value reaches 1.12 Mbps for $BER=0$ and $n=1$. Along with an increasing value of BER the presented curves flatten out. For a given BER, the decrease of $S_{DATA}$ together with an increase of $n$ is related to a growing number of collisions in the medium. The observed decline in the value of $S_{DATA}$ between $BER=0$ and $BER=10^{-5}$ is very small.

Fig. 4 presents $S_{DATA}$ as a function of $n$ for different values of frame length and $BER=0$. For a given $n$, an increasing frame length leads to a fall in the attainable $S_{DATA}$.

Fig. 5 represents the correlation between $S_{DATA}$ and $n$, for different values of frame length and $BER=10^{-5}$, while Fig. 6 displays $S_{DATA}$ as a function of $n$ for different frame lengths and $BER=10^{-4}$. Compared to the values obtained for $BER=0$, we observe a reduction in the value of $S_{DATA}$ due to the influence of channel errors.

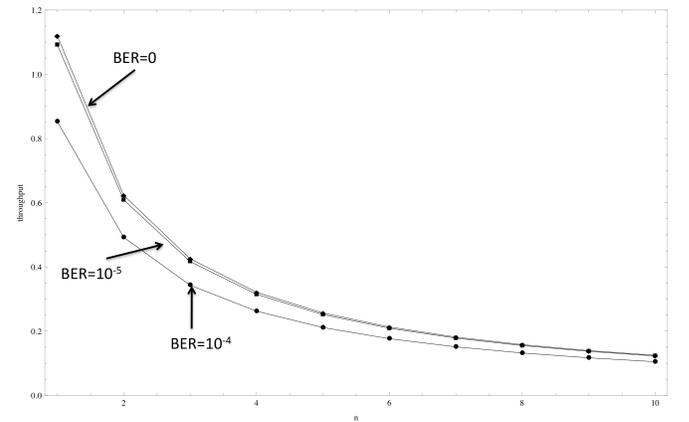

Fig. 3. $S_{DATA}$ as a function of $n$ – for $L=214$ octets and different values of BER

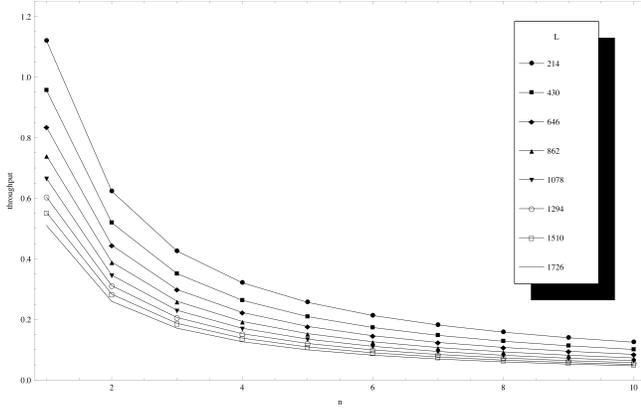

Fig. 4. $S_{DATA}$ as a function of $n$ – for different values of frame length and $BER$=0

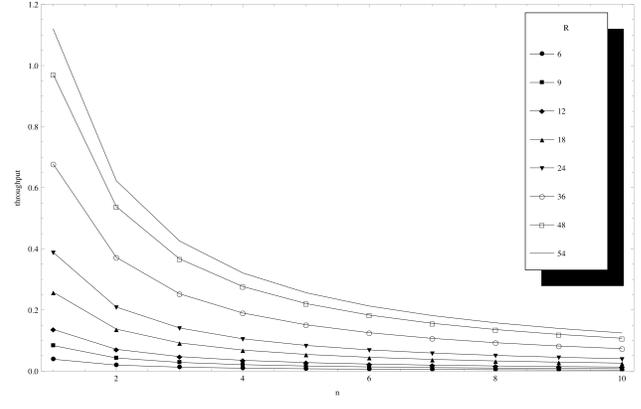

Fig. 7. $S_{DATA}$ as a function of $n$ – for $L$=214 octets, $BER$=0 and different values of $R$

## B. ACK frames

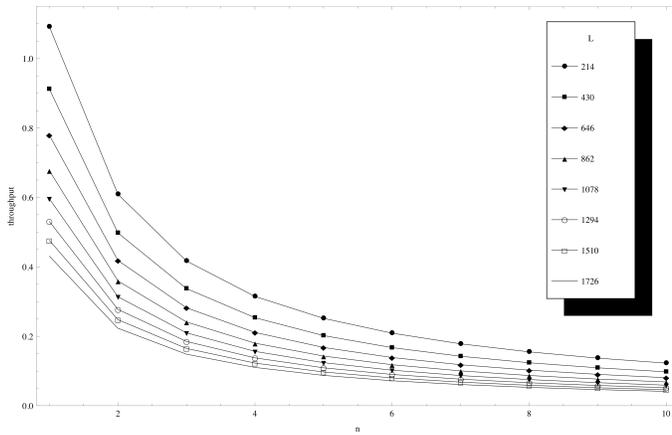

Fig. 5. $S_{DATA}$ as a function of $n$ – for different values of frame length and $BER$=$10^{-5}$

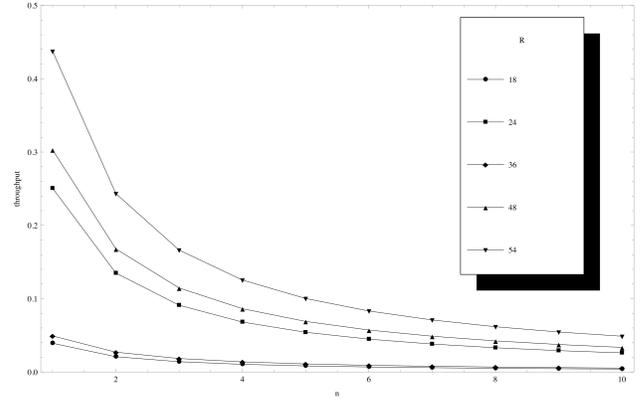

Fig. 8. $S_{ACK}$ as a function of $n$ – for $L$=214 octets, $BER$=0 and different values of $R$

We evaluate (Fig. 8) $S_{ACK}$ as a function of $n$ for different IEEE 802.11g data rates $R\in\{18, 24, 36, 48, 54\}$ Mbps. For n=1 and R=54, $S_{ACK}$=0.44 Mbps (82 bits serve as a hidden channel). The throughput for 24 Mbps networks is higher than for 36 Mbps because of the different capacity of the hidden channel: 58 and 10 bits respectively.

## VI. CONCLUSIONS AND FUTURE WORK

In this paper we evaluated a new steganographic method called WiPad intended for IEEE 802.11 OFDM networks, whose functioning bases on insertion of bits into the padding of transmission symbols. The analysis for IEEE 802.11g 54 Mbps networks revealed that the capacity of WiPad equals 1.1 Mbit/s for data frames and 0.44 Mbit/s for ACK frames, which gives a total of almost 1.65 Mbit/s. To the authors' best knowledge this is the most capacious of all the known steganographic network channels.

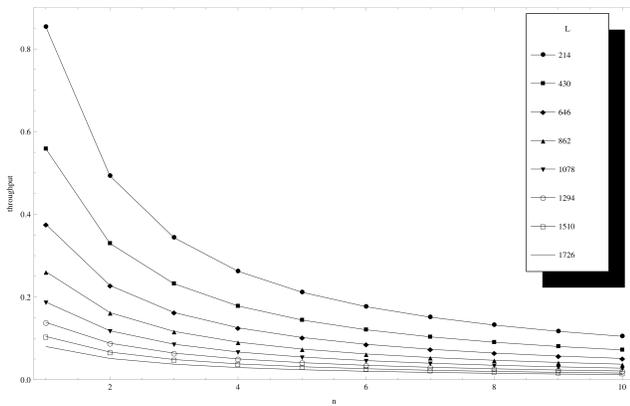

Fig. 6. $S_{DATA}$ as a function of $n$ – for different values of frame length and $BER$=$10^{-4}$

Finally we evaluate (Fig. 7) $S_{DATA}$ as a function of $n$ for different IEEE 802.11g data rates $R\in\{6, 9, 12, 18, 24, 36, 48, 54\}$ Mbps.

Future work will include WiPad analysis for typical IP packet sizes and the estimation of achievable steganographic bandwidth in case of the IEEE 802.11n standard. Further studies should also involve pinpointing potential detection mechanisms of the proposed communication system.